\documentclass[aps,pra,reprint,superscriptaddress,longbibliography,amsmath,amssymb]{revtex4-1}
\usepackage[utf8]{inputenc}
\usepackage{amsfonts}
\usepackage{physics}
\usepackage{graphicx}
\usepackage{overpic}
\usepackage{color}
\usepackage[dvipsnames]{xcolor}
\usepackage[hidelinks]{hyperref}
\usepackage{comment}
\usepackage{tikz}
\usepackage[T1]{fontenc} 
\usepackage{enumitem}
\usepackage{booktabs}
\usepackage{float}
\usepackage{amsthm}

\usepackage{algorithm}
\usepackage{algpseudocode}
\usepackage{soul}

\newcommand{\unipd}{Dipartimento di Scienze Chimiche,
 Universit{\`a} degli Studi di Padova, Italy I-35131, Padova, Italy}
\newcommand{\padcen}{Padua Quantum Technologies Research Center,
  Universit{\`a} degli Studi di Padova}

\newtheorem*{theorem}{Compressed sensing off-the-grid}
\newtheorem*{lemma}{Frequency detectability condition}
\newtheorem*{lemma2}{Number of detectable frequencies}
\newtheorem*{lemma3}{Error bound on detectable frequencies}

\begin{document}

\title{Heisenberg limited multiple eigenvalue estimation via off-the-grid compressed sensing}


\author{Davide Castaldo}
\email{davide.castaldo@phys.chem.ethz.ch}
\affiliation{\unipd}
\affiliation{Department of Chemistry and Applied Biosciences, ETH Zurich, Vladimir-Prelog-Weg 2, 8093 Zurich, Switzerland}

\author{Stefano Corni}
\affiliation{\unipd}
\affiliation{\padcen}
\affiliation{Istituto Nanoscienze—CNR, via Campi 213/A, 41125 Modena (Italy)}

\begin{abstract}

Quantum phase estimation is the flagship algorithm for quantum simulation on fault-tolerant quantum computers. We demonstrate that an \emph{off-grid} compressed sensing protocol, combined with a state-of-the-art signal classification method, enables the simultaneous estimation of multiple eigenvalues of a unitary matrix using the Hadamard test while sampling only a few percent of the full autocorrelation function. Our numerical evidence indicates that the proposed algorithm achieves the Heisenberg limit in both strongly and weakly correlated regimes and requires  very short evolution times to obtain an $\epsilon$-accurate estimate of multiple eigenvalues at once.

Additionally---and of independent interest---we develop a modified off-grid protocol that leverages prior knowledge of the underlying signal for faster and more accurate recovery. Finally, we argue that this algorithm may offer a potential quantum advantage by analyzing its resilience with respect to the quality of the initial input state.
\end{abstract}

\maketitle

\section*{Introduction}

As quantum computing continues to advance, the challenge of extracting practical benefits from this technology becomes increasingly daunting. Since the initial intuition that quantum computers could offer computational advantages in simulating chemistry and materials \cite{feynman2018simulating, aspuru2005simulated, reiher2017elucidating}, our understanding of the full quantum computing stack---from physical implementation to high-level algorithms---has evolved significantly\cite{briegel2009measurement, bartolucci2023fusion, aharonov2008adiabatic}.

On the hardware side, we are on the verge of developing the first prototypes of scalable, error-correctable architectures \cite{krinner2022realizing, google2023suppressing, putterman2025hardware, acharya2024quantum, sivak2023real}. However, these advances come with the realization that near-term machines may struggle to maintain the quality of individual physical qubits as architectures scale up \cite{katabarwa2024early}.

This indication, combined with recent resource estimates for the most promising implementations of the quantum phase estimation (QPE) algorithm\cite{dalzell2023quantum}---widely regarded as the flagship algorithm for simulating quantum systems \cite{rubin2023fault,castaldo2024differentiable, loaiza2024nonlinear, steudtner2023fault}---suggests that millions of gates and hundreds of logical qubits would be required to surpass current state-of-the-art classical simulation techniques. These estimates stem from three key factors: (i) the standard textbook QPE algorithm \cite{cleve1998quantum} requires high connectivity between all qubits and it is extremely sensitive to noise\,\cite{nelson2024assessment}, as the readout and qubit registers become highly entangled due to multiple conditioned evolutions of the system register on different qubits of the readout register; (ii) for systems of interest, it is often necessary to prepare quantum states with high overlap with the target state \cite{erakovic2025high, fomichev2024initial, ollitrault2024enhancing}; and (iii) the most promising routines for performing time evolution, while asymptotically efficient \cite{berry2014exponential, berry2015simulating, low2019hamiltonian, somma2025shadow, luo2025efficient, kiumi2024te}, still produce prohibitively deep circuits.\\

Consequently, the development of QPE-like algorithms aimed at bringing  a practical quantum advantage closer to reality\,\cite{beverland2022assessing} is of utmost current interest. All these algorithms share the common idea that QPE allows one to estimate (some of) the eigenvalues of a unitary $ U $ by sampling and Fourier transforming the autocorrelation function of a quantum Hamiltonian $H$. Bearing this in mind, the ultimate goal is to simultaneously reduce (i) the circuit width, (ii) the number of circuit executions, and (iii) the circuit depth to achieve an given error $ \epsilon $ in the estimate of a single (or multiple) eigenvalue(s). This cost reduction is achieved by exploiting the analogy between the QPE algorithm and a signal recovery problem. We will draw this parallelism in more detail in the next section; here, we briefly recall the main works that are related to this study\cite{ding2023even, ding2024quantum, yi2024quantum, ni2023low}. \\

First, we cite the work of Yi \textit{et al.}~\cite{yi2024quantum}, which inspired this study. Indeed, the authors of Ref.~\cite{yi2024quantum} were the first to propose using compressed sensing to develop an algorithm suitable for EFTQCs (Early Fault-Tolerant Quantum Computer) for estimating a single eigenvalue. Compressed sensing is a signal processing technique that enables accurate reconstruction of sparse signals from far fewer samples than required by the Whittaker-Nyquist-Shannon theorem\cite{whittaker1915xviii}. The method relies on two key principles: (1) \emph{sparsity} - the signal must have a concise representation in some basis (e.g., most of its components are zero), and (2) \emph{incoherence} - the sampling strategy must capture the signal's information content. Remarkably, this allows perfect recovery of $s$-sparse $N$-dimensional signals using only $\mathcal{O}(s \log N)$ measurements \cite{candes2006compressive, donoho2006compressed}. In quantum phase estimation, compressed sensing proves particularly valuable as the spectral decomposition of quantum states often exhibits natural sparsity in the frequency domain.

Thus, using this technique the authors of Ref.\,\cite{yi2024quantum} have demonstrated that a quantum computer can be used to sample a correlation function containing a \emph{single} dominant frequency and estimate this frequency with very few samples by reconstructing it in the frequency domain.

Here we extend their results in three key aspects: (i) we develop an efficient and optimal algorithm (i.e., saturating the Heisenberg limit\cite{giovannetti2006quantum}) for \emph{multiple} eigenvalue estimation; (ii) we demonstrate our algorithm's resilience to poor initial-state preparation on a prototypical strongly correlated molecular system; and (iii) we introduce a specialized off-the-grid recovery algorithm that is both classically efficient and robust to shot noise in realistic scenarios. The last point is particularly important because makes our proposal practical from a computational point of view as, commonly, off-grid methods avoid gridding errors at the cost of the computational overhead of solving a non-linear optimization problem \cite{tang2013compressed}. Another work relevant to this study is that of Ding \textit{et al.}\cite{ding2024quantum} which, as noted in \cite{yi2024quantum}, is also closely related to a compressed sensing protocol named Orthogonal Matching Pursuit\,\cite{cai2011orthogonal} and targets the problem of multiple eigenvalue estimation in an EFTQC scenario using few samples and short evolution times.

Both these two previous works and the present one, benefited from several other contributions. Particularly, Ref.\cite{somma2019quantum} recognized the connection between the signal recovery problem and the QPE (single or multiple) task. Furthermore, in Refs.\,\cite{ding2023even, li2023adaptive}, the authors achieved algorithms whose minimal required maximal time evolution $T_{max}$ was bounded only by the minimal gap $\Delta_f$ of the dominant frequencies (eigenvalues) in the signal. As we shall see later, this is a feature that our algorithm directly inherits from the properties of off-grid protocols\cite{tang2013compressed}. 

Finally, we also mention the work of Kick\,\cite{kick2024super} \textit{et al.} which very recently applied a superresolution approach to the problem of large scale simulation of electronic spectra at the TDDFT level of theory.

This paper is organized as follows. In Sec.\,\ref{theory} we formulate the eigenvalue estimation problem as a signal recovery problem. Building on this, we present the main result of this work: the development of a gridless recovery algorithm to estimate autocorrelation functions from a few samples. Since our recovery algorithm works in the time domain, this subroutine is coupled with a well-known subspace method that extracts frequencies in multicomponent noisy signals\,\cite{liao2016music}. 
In Sec.\,\ref{numerics} we provide numerical evidences for the implementation of the algorithm and discuss its computational scaling. Particularly, in Sec.\,\ref{heisenberg_scaling}, we show that we are able to recover simultaneously ground and excited state energies at Heisenberg scaling. Here we also discuss the effect of the signal length on our frequency resolution capability and accuracy estimate: we anticipate that the algorithm outperforms the standard QPE scaling\cite{nielsen2010quantum} and is on par with state-of-the-art methods\cite{ding2024quantum} as $T_{max}\pi \ll \epsilon^{-1}$. Finally, we study the accuracy of the estimation protocol in estimating the fourth excited singlet state energy of the $H_8$ hydrogen chain as a function of the initial state input overlap with the exact FCI wavefunction. As previously mentioned, this is also an important factor to consider when developing a quantum algorithm which aims at demonstrating quantum advantage as it requires achieving chemically accurate results for systems for which we may only upload on the quantum computer approximate initial states. The conclusions include some remarks on potential improvements of this work and possible future directions for the field of EFTQC algorithms for quantum chemistry.


\section{Theory}\label{theory}
\subsection{Molecular excitation energies as a signal recovery problem}\label{sec:MPS}

In this section, we formulate the problem of estimating multiple eigenvalues of a quantum Hamiltonian, with particular attention to the special case of the estimation of molecular excitation energies, as a signal recovery problem.

This equivalence is simply found by noting that the autocorrelation function (expressed in the eigenbasis) is a complex signal made up of a mixture of complex exponentials spanning a frequency bandwidth $\mathcal{B}$:

\begin{equation}
    z(t) = \langle \psi(0)|\psi(t) \rangle = \sum_i^{|\mathcal{E}|} |c_i|^2 e^{-i E_i t}
    \label{autocorr_signal}
\end{equation}
where $|\mathcal{E}|$ is the number of eigenstates of the Hamiltonian $H$ of the system with non-zero overlap with the initial state $|\psi(0)\rangle$. Moreover, we denote $E_i$ the energy of the i-th eigenstate and $|c_i|^2 = |\langle \psi(0) | i \rangle |^2$.

The question we try to answer in this work is the following: is it possible to recover a relevant subset of frequencies $|\mathcal{\Tilde{E}}|$ from the complete $|\mathcal{E}|$ acquiring a few noisy samples from $z(t)$? This question, even though easily formulated, requires several specifications: if so, what is the total quantum runtime of the routine? What do we mean by few measurements? What do we mean by noisy? How large can be the subset $|\mathcal{\Tilde{E}}|$ w.r.t. to $|\mathcal{E}|$? What is the maximal evolution time $T_{max}$ at which we need to sample our signal?

\begin{figure}
    \centering
    \includegraphics[width=\linewidth]{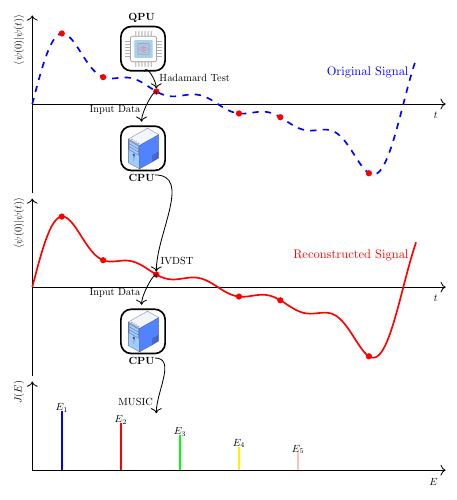}
    \caption{Schematic representation of the proposed protocol. Samples of the autocorrelation function are measured on a quantum processing unit via the Hadamard test, see Sec.\,\ref{sec:MPS}. The collected data are used to feed a signal recovery algorithm,  see Sec.\,\ref{recovery_section}. Finally, the reconstructed noisy signal is used to accurately estimate its dominant components, see Sec.\,\ref{denoising}.}
    \label{algorithm_summary}
\end{figure}

These questions concern critical quantities that characterize the signal recovery problem and hold significance for the practical implementation of the quantum algorithm. We will try to address them in detail in this work.

First, we define the total quantum runtime as 

\begin{equation}\label{runtime}
T_{runtime} = M \sum_n |t_n|    
\end{equation}
where $t_n$ are the time instants at which we sample the autocorrelation function and $M$ is the number of shots per sample. This definition has been adopted in previous related works to estimate the efficiency of the proposed method\cite{yi2024quantum, ding2024quantum, ding2023even, ding2023simultaneous}. We note that, for this kind of metric, the scaling of the routine with the size of the system studied is implicit as the cost of obtaining the autocorrelation function for the same value of $t_n$ is different for systems of different sizes.

In the following we will denote the sampling noise as $\xi(\eta)$ with $\eta$ being its variance.

The noisy samples of the autocorrelation function are defined as:

\begin{equation}
    \Tilde{z}(t_n) = \mathbb{E} [ \langle \psi(0)|U(0, t_n)| \psi(0)\rangle ]_{M}
\end{equation} 
where $\mathbb{E}[\cdot]_M$ is the average over $M$ independent shots. Moreover, we recall that the sampling noise variance and the number of samples are related by the expression $\eta = \frac{1}{M}$ due to the central limit theorem. Subsequently we have that:

\begin{equation} \label{noise_per_point}
   \Tilde{z}(t_n)   = z(t_n) + \xi(\eta) 
\end{equation}
It is important to notice that in this work we focus on the properties of the autocorrelation function and its relation with the estimated energies. The error due to the quantum simulation subroutine, $\epsilon_{sim} = \mathcal{O}(||H|| t_n)$, will not be considered in our analysis as it may be reduced arbitrarily increasing the depth of the circuit (or even completely vanish assuming access to a block encoding of $H$\cite{low2019hamiltonian}) and depends on the particular quantum simulation algorithm implemented, thus going beyond the scope of this work.

Here, we focus on developing an algorithm that recovers only a subset of the energy spectrum of $H$. Accordingly, we decompose the autocorrelation function into contributions from different parts of the spectrum:

\begin{equation}
    z(t) = \sum_l^{|\mathcal{\Tilde{E}}|}|c_l|^2 e^{- i E_l t} + \sum_{i \neq l}^{|\mathcal{E}|-|\mathcal{\Tilde{E}}|} |c_i|^2 e^{- i E_i t} = z_r(t) + z_{nr}(t)
\end{equation}
where the last identity defines the relevant part and \textit{not} relevant part of the autocorrelation function $z_r(t)$ and $z_{nr}(t)$, respectively. Finally, we add to Eq.\,\ref{noise_per_point} the contribution of the non relevant part as part of the noise:

\begin{equation}
\label{relevant_not_relevant}
    \tilde{z}(t_n) = z_r(t_n) + z_{nr}(t_n) + \xi(\eta) 
\end{equation}

To conclude this section, we recall that noise can also be characterized in terms of additional spurious frequency components appearing in the signal. Depending on the stochastic process that generates them, these new components have different frequency, amplitude, and phase distributions. When it comes to shot-noise measurements, the frequency distribution is flat throughout the signal frequency bandwidth (i.e., $\omega_k \in \mathcal{B}$), the amplitudes correspond to the noise standard deviation and ensure normalization and the phases $\phi_k$ are uniformly distributed in the interval $[0, 2\pi]$:

\begin{equation}
\label{noisy_signal}
    \Tilde{z}(t_n) = z_r(t_n) + \underbrace{z_{nr}(t_n) + \frac{1}{\sqrt{M}}\sum_k c_k e^{-i (\omega_k + \phi_k)t}}_{z_{noise}(t_n)}
\end{equation}

We have now defined all the necessary quantities to discuss the eigenvalue problem in terms of a signal recovery problem. In the following pages we lay down a recovery-denoising algorithm able to recover the frequencies of $z_r(t)$ from few noisy samples of $\Tilde{z}(t)$. In Sec.\,\ref{recovery_section} we focus on describing a protocol proposed in Ref.\,\cite{wang2018ivdst} to recover a classical signal in time domain which we modified to suit measurements coming from a quantum computer. Subsequently, in Sec.\,\ref{denoising} we present the MUltiple SIgnal Classification (MUSIC) algorithm\cite{liao2016music} used to denoise and transform the reconstructed signal into the frequency domain.

\subsection{Off-the-grid compressed sensing}\label{recovery_section}

First, we introduce the main results of Ref.\,\cite{tang2013compressed} which provides the foundational theorems for the off-the-grid sensing protocol.

We recall the main ideas behind compressed sensing: supposing that we have an N-dimensional signal which is $s$-sparse in some basis (i.e., it has only $s$ nonzero elements in that basis), we want to recover the full signal taking only $m \approx \mathcal{O}(s) \ll \mathcal{O}(N)$ samples using the \textit{a priori} knowledge of its sparsity. 
The following theorem (which we only state), relates the sample complexity of the algorithm to guarantees on the recovery protocol:

\begin{theorem}
    Suppose we have a signal such as Eq. \ref{autocorr_signal} with minimal frequency gap $\Delta_f = \min_{i, j} |E_i - E_j| \geq \frac{1}{N}$, then we can recover the overall signal with a semidefinite program and localize its frequencies with probability at least $1 - \delta$ acquiring $m$ samples at random in time domain, with $m$ given by:

    \begin{equation}
    \label{number_of_samples_eq}
        m \geq C \max \big \{ \text{log}^2\frac{N}{\delta}, s \text{log} \frac{s}{\delta} \text{log}\frac{N}{\delta} \big \} 
    \end{equation}
    where $C$ is a positive, problem dependent constant.
\end{theorem}

The last theorem provides a lower bound on the number of samples needed to reconstruct the signal. The question on how to reconstruct it is still unanswered. Thus, the authors define a semidefinite program to efficiently compute $z(t_n)$ from $m$ random samples of $\Tilde{z}(t_n)$. For sake of completeness, we refer the reader to Ref.\,\cite{wolkowicz2012handbook} for a detailed account on semidefinite programming. \\
 
 We call $\mathcal{S}$ the set of $m$ randomly sampled points, $x$ the solution vector, $\cdot _{\mathcal{S}}$ is any vector whose space is restricted to the subspace of sampled points. Such a notation can be also replaced by introducing a linear compressive operator $P \in \mathbb{C}^{m\, \text{x}\, N}$ such that $z_{\mathcal{S}}= Pz$. For our purposes $P$ will be always obtained discarding from the identity matrix the rows corresponding to the not measured samples. In order to reconstruct the autocorrelation function we aim to solve:

\begin{equation}
\label{atomic-norm-minim}
\begin{aligned}
    &\min_{x} ||x||_{\mathcal{A}} \\
    &\text{s.t.} \quad 
\|x_{\mathcal{S}} - z_{\mathcal{S}}\|_2
    \leq \gamma
\end{aligned}
\end{equation}

Where the objective function of the minimization problem is the atomic norm defined as:

\begin{equation}
\|x\|_{\mathcal{A}} = \inf \left\{\sum_{k} c_k : x = \sum_{k} c_k \phi(\omega_k), \, \omega_k \in [0, 1), \, c_k \geq 0 \right\}
\end{equation}
and $\phi(\omega_k) = e^{i \omega_k t}$ are usually referred to as \textit{atoms} of the optimization.

The solution to Eq.\,\ref{atomic-norm-minim} is very hard and computationally demanding. For this reason, the authors have reformulated Eq.\,\ref{atomic-norm-minim} as a semidefinite program which can be efficiently solved (i.e., in polynomial time): 

\begin{equation}
\label{beurling-lasso}
\begin{aligned}
    &\min_{u, x, t} \lambda \left( \text{Tr}\{\text{T}(u)\} + \frac{1}{2}t \right) + \|x_{\mathcal{S}} - z_{\mathcal{S}}\|_2 \\
    &\text{s.t.} \quad 
    Z(u, x, t) = \begin{pmatrix}
        \text{T}(u) & x \\
        x^{\dagger} & t
    \end{pmatrix}
    \geq 0
\end{aligned}
\end{equation}
Where the vector $u$ is an auxiliary vector determined by the same atomic set that defines $x$ and $\lambda$ is a regularization parameter. Moreover, $T(u)$ is defined as the Toeplitz matrix associated to the $u$ vector:

\begin{equation}
    T(u) =
\begin{bmatrix}
u_0 & u_1 & u_2 & \cdots & u_{n-1} \\
u_{-1} & u_0 & u_1 & \cdots & u_{n-2} \\
u_{-2} & u_{-1} & u_0 & \cdots & u_{n-3} \\
\vdots & \vdots & \vdots & \ddots & \vdots \\
u_{- (n-1)} & u_{- (n-2)} & u_{- (n-3)} & \cdots & u_0
\end{bmatrix} ,
\end{equation}
please notice that in the following we will refer to signals defined in the time interval [-T, T]. Analogously, negative indexes as in the previous equation refer to points of the vectorb belonging to the interval [-T, 0].

Despite the fact that the solution can be achieved efficiently\,\cite{toh1999sdpt3}, conventional convex solvers would still require a significant amount of time to recover the true signal. This has motivated researchers in the field of signal processing to develop algorithms to recover the underlying signal faster and more accurately.\\

The first result of this work is an adaptation of the Iterative Vandermonde Decomposition and Shrinkage-Thresholding (IVDST) algorithm developed in Ref.\,\cite{wang2018ivdst} to recover autocorrelation functions of quantum Hamiltonians.

The main idea of Ref.\,\cite{wang2018ivdst} is to rewrite Eq.\,\ref{beurling-lasso} as an unconstrained optimization problem:

\begin{equation}\label{unconstrained_opt}
\min_{u, x, t} F(u, x, t) =  
\frac{\mu}{2} \|x_{\mathcal{S}} - z_{\mathcal{S}}\|_2
+ \mathrm{tr}(Z(u, x, t)) + C(Z(u, x, t))
\end{equation}
where $Z(u, x, t)$ has been defined in Eq.\,\ref{beurling-lasso} and $C(Z(u, x, t))$ is a function which takes the value of 0 if $Z(u, x, t)$ is positive semidefinite and hermitian or takes the value of $\infty$ if otherwise. This behavior renders the cost function non-differentiable, prompting the authors of Ref.\,\cite{wang2018ivdst} to adopt an iterative approach as an efficient solution to Eq.\,\ref{unconstrained_opt}.

\begin{algorithm}[H]
\caption{Modified IVDST}
\begin{algorithmic}[1]
\State \textbf{PII}($\{\omega_{guess} \}$) = $\theta:  (T_1, x_1, t_1) = (T_0, x_0, t_0)$
\While{Stopping criterion not met}
    \State \textbf{Smoothing:} Set $\bar{\theta}_i$ by:
    \[
    \bar{\theta}_i = \theta_i + \frac{t_{i-1} - 1}{t_i} (\theta_i - \theta_{i-1}), \quad \bar{t}_i = \frac{1 + \sqrt{4t_{i-1}^2 + 1}}{2}, \, t_0 = 1
    \]
    \State \textbf{Gradient descent:} Update $\theta_g$ by:
    \[
    x_g = \bar{x}_i - \delta P^\dagger (P\bar{x}_i - z_{\mathcal{S}}), \, t_g = \bar{t}_i, \, T_g = \bar{T}_i
    \]
    \State  \textbf{Singular Value Decomposition:} $T_g = SD V^\dagger$
    \State   \textbf{Shrinkage-thresholding:}
    \[
    \tilde{D} = \mathrm{diag} (\max(\mathrm{diag}(D) - \tau \mathbf{1}, 0))
    \]
    \[
    \tilde{T} = S\tilde{D}V^\dagger
    \]
    \State  \textbf{PSD conditioning:} Compute $\tilde{Z}$ by:
    \[
    \tilde{Z} = U_S \Sigma_S U_S^\dagger, \quad (\Sigma_S, U_S) = \text{eigs}(Z, S)
    \]
     \qquad \quad where $Z = \begin{bmatrix} t_g & x_g^\dagger \\ x_g & \tilde{T} \end{bmatrix}$ and $S = \mathrm{rank}(\tilde{D}) + 1$
    \State \textbf{Update:}
    \[
    x_{i+1} = \tilde{Z}_{2:N+1,1}, \, t_{i+1} = \tilde{Z}_{1,1}, \, T_{i+1} = \mathcal{T}(\tilde{Z}_{2:N+1, 2:N+1})
    \]
    \State \textbf{Stopping criterion:} Stop if $\frac{\|\tilde{T}_{i+1} - T_i\|_F}{\|T_i\|_F} \leq \gamma$
\EndWhile
\end{algorithmic}
\label{IVDST}
\end{algorithm}

In this work we adopted a modification to their original proposal which we report as a pseudocode (see Alg. \ref{IVDST}). Our modification makes the recovery algorithm both (i) compatible with noisy data coming from a quantum computer and (ii) faster due to a physically informed preconditioning of the initial guess.

First, we consider our modification that enables the IVDST algorithm to handle noisy measurements. In particular, in the original article to enforce a low-rank (that is, sparse) solution, the authors propose to use a Vandermonde decomposition\,\cite{hua1990matrix} (hence the name of the algorithm) of the matrix $T_g = VDV^{\dagger}$. We recall that a Vandermonde decomposition is a specific form of matrix  factorization that allows to represent them in diagonal form ($D$ is, indeed, diagonal) via transformation matrices that have a Vandermonde structure. Their motivation is in the computational efficiency of this routine which is less demanding than a singular value decomposition. This is the optimal solution when it comes to the reconstruction of a noiseless signal. Nevertheless, we found that including noisy samples coming from quantum measurements, the Vandermonde decomposition becomes really unstable, preventing the algorithm to converge to the desired solution. As we show in Sec.\,\ref{numerics} by substituting the Vandermonde decomposition with a Singular Value Decomposition (step 5 of Alg.\ref{IVDST}) we are able to accurately recover the signals with a number of shots per sample that scales as $ \#_{shots} = \mathcal{O}(\sqrt{s \, log s \, N\, log N }) $ where $N$ is the total length of the signal and $s$ is the number of dominant frequencies in the signal.

\begin{algorithm}[H]
\caption{Physically Informed Initialization (PII)}
\begin{algorithmic}[1]
\State \textbf{Input}: $\{\bar{c}_{guess}, \bar{\omega}_{guess} \}$
    \State Initialize $x_0$:
    $x_0 = \sum_{c, \omega \in \{ \bar{c}_{guess}, \bar{\omega}_{guess} \} } c\, e^{- i \omega t}$
    \State Initialize $T_0$: $T_0 = \mathcal{T}(x_0x_0^{T})$
    \State Initialize $t_0$: $t_0 =  \frac{Tr(T_0)}{N}$
\State \textbf{Output}: $\theta$ : $(x_0, T_0, t_0)$
\end{algorithmic}
\label{initialization}
\end{algorithm}

Secondly, we modified the guess initialization routine to account for previous knowledge of the system studied. We have summarized the original initialization routine and our proposed initialization in Alg.\ref{og_initialization} and Alg.\ref{initialization}, respectively. In particular, instead of using the initial measurements to generate the initial input guess (Alg.\ref{og_initialization}), we now use a synthetic signal specified by frequencies ${\bar{\omega}_{guess}}$ and coefficients ${\bar{c}_{guess}}$ that come from cheaper classical calculations.

We point out that the operator $\mathcal{T}(\cdot)$ returns a Toeplitz matrix by averaging all elements along each diagonal direction.

\begin{algorithm}[H]
\caption{IVDST initialization\cite{wang2018ivdst}}
\begin{algorithmic}[1]
\State \textbf{Input}: $z_{\mathcal{S}}$
    \State Initialize $x_0$:
    $x_0 = z_{\mathcal{S}}$
    \State Initialize $T_0$: $T_0 = \mathcal{T}(z_{\mathcal{S}}z_{\mathcal{S}}^{T})$
    \State Initialize $t_0$: $t_0 =  \frac{Tr(T_0)}{N}$
\State \textbf{Output}: $\theta$ : $(x_0, T_0, t_0)$
\end{algorithmic}
\label{og_initialization}
\end{algorithm}

\begin{figure}
    \centering
    \includegraphics[width=\linewidth]{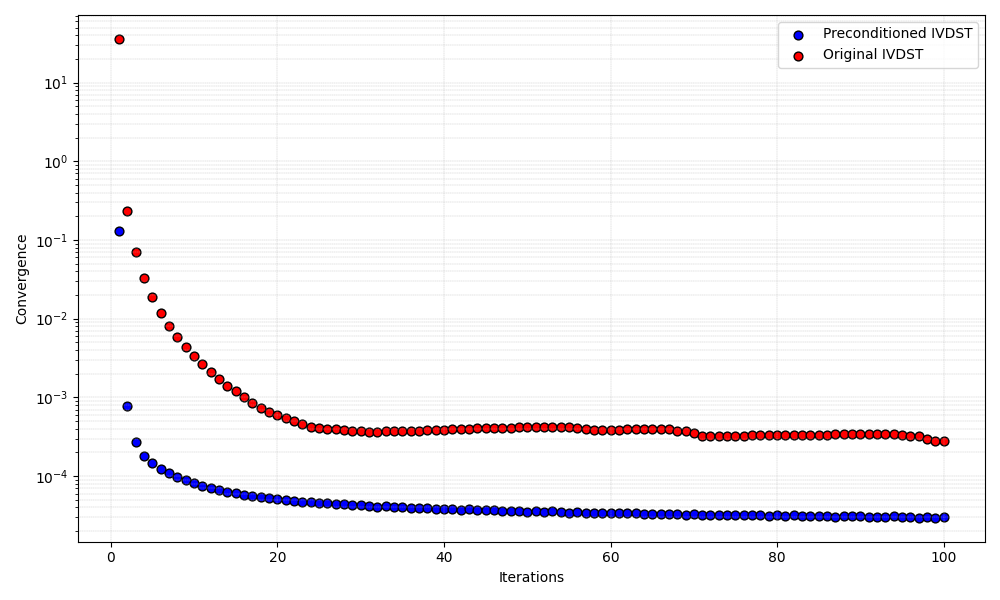}
    \caption{Preconditioning the IVDST with approximate solutions allows for faster signal reconstruction. Red dots show the convergence of the recovery algorithm using the initialization strategy reported in Alg.\,\ref{og_initialization} while blue dots correspond to the convergence of the recovery algorithm using the initialization proposed in this work (see Alg.\,\ref{initialization}).}
    \label{convergence_plot}
\end{figure}

This modification improves the original proposal in three ways: (i) increases the accuracy of the signal recovery, (ii) provides orders-of-magnitude speed-ups in the convergence of the optimization, and (iii) allows the integration of this algorithms with classical quantum chemistry methods which can now provide a warm-started guess to the optimization. We note that such a modification does not apply to the specific case of quantum autocorrelation functions but could be of use in any other sector where the IVDST algorithm is of use. In Fig.\,\ref{convergence_plot} we have reported convergence in the first 100 iterations of the IVDST algorithm with (Alg.\ref{initialization}) or without (Alg.\ref{og_initialization}) preconditioning the recovery protocol with a signal that uses prior knowledge of the seeked solution. A typical scenario, relevant to the numerical experiments in this work, involves using a low-cost computational method (such as CISD or CASCI on a smaller active space) to estimate the frequencies of the target states. The coefficients for each component can then be set based on an estimated overlap with the true state or, as done in this work, by simply assigning equal coefficients to all components. As we can see, preconditioning the recovery algorithm with an initial guess close to the underlying signal allows to dramatically speed up the convergence. In Fig.\,8 of Sec.\,\ref{preconditiong_results} we also report the results obtained on the LiH molecule (as studied in Fig.\,\ref{heinsenb-scaling1}) when estimating four different eigenvalues of an autocorrelation function recovered with (or without) the initial physically informed guess of Alg.\,\ref{initialization}. We found that, consistently, the recovery with a physically informed initial guess can achieve better accuracy with the same quantum computational budget.

\subsection{From time to frequency domain: denoising with MUSIC}\label{denoising}

In the previous section we have elucidated how to recover the autocorrelation signal from few measurements via an off-the-grid compressed sensing approach.

From now on, we will consider the solution $x^{*}(t)$ of Eq.\,\ref{beurling-lasso} as a proxy for the original signal $z(t)$. 

In order to estimate part of the spectrum of the Hamiltonian $H$ we still need to discuss how to recover frequencies out of the time-domain signal. The first, natural, option is to perform a Fourier transform. Unfortunately, this approach has been shown to lead to a shot-noise-limited accuracy scaling of the protocol (i.e. $T_{runtime} = \mathcal{O}(\epsilon^{-2})$)\,\cite{dutkiewicz2022heisenberg}. We can rationalize this scaling looking at Eq.\,\ref{noisy_signal} where frequencies associated to the presence of noise appear with amplitude given by the inverse square-root of the number of shots. In Appendix A (see Fig.\,\ref{fft2music}) we compare the spectrum obtained with a straightforward Fourier transformation to the spectrum obtained with the MUSIC algorithm used in this work, and developed by Liao and Fannjiang\cite{liao2016music}, to denoise the signal and obtain energy estimates with Heisenberg scaling (see Sec.\,\ref{numerics}). As one can see, the accuracy achieved in the frequency estimation for the same signal is two order of magnitude higher using the MUSIC spectrum rather than the FFT spectrum (see Tab.\,\ref{tab:fft_music}). 

The algorithm is able to denoise and identify the most prominent frequencies in the signal by decomposing the signal space $\mathcal{F}$ into two orthogonal subspaces $\mathcal{F} = \mathcal{F}_s \oplus \mathcal{F}_n$ in which the components of the signal and the noise are spanned, respectively.

The first step consists of building a Hankel matrix from the recovered signal:

\begin{equation}
    \textbf{W} = \text{Hank}(x^*(t))
\end{equation}
where $\text{Hank}(x^*(t))$ is defined as:

\begin{equation}
   \textbf{W} =  \begin{bmatrix}
x_0 & x_1 & x_2 & \cdots & x_{n-1} \\
x_1 & x_2 & x_3 & \cdots & x_n \\
x_2 & x_3 & x_4 & \cdots & x_{n+1} \\
\vdots & \vdots & \vdots & \ddots & \vdots \\
x_{n-1} & x_n & x_{n+1} & \cdots & x_{2n-2}
\end{bmatrix}
\end{equation}

We recall that Hankel matrices appear in other signal recovery techniques such as the Prony's method and the pencil method\,\cite{sarkar1995using} as they encode correlations between the signal at different points in time.

Subsequently, the two subspaces are identified performing a singular value decomposition of the matrix $\textbf{W}$:

\begin{equation}
    \textbf{u}, \textbf{v}, \textbf{D} = SVD(\textbf{W})
\end{equation}

Setting the dimension of the signal subspace $\mathcal{F}_s$ equal to the number of frequencies $s$ one wants to detect ($s = |\mathcal{\Tilde{E}}|$), it is possible to build the projector onto the noise subspace $\textbf{P}_n$ as:

\begin{equation}
   \textbf{P}_N =  \mathbb{I} - \sum_i^{|\mathcal{\Tilde{E}}|} |u_i\rangle \langle u_i |
\end{equation}

Finally, to identify the frequencies belonging to $\mathcal{F}_s$, we look for the maxima of the imaging function $J(\omega)$:

\begin{equation}
\label{imaging_function_eq}
    E_i = \text{argmax}(J(\omega)) = \text{argmax}(\frac{\langle a(\omega) | a(\omega) \rangle }{\langle a(\omega) | \textbf{P}_N^{\dagger}\textbf{P}_N | a(\omega) \rangle })
\end{equation}
where $|a(\omega)\rangle$ is defined as $|a(\omega)\rangle = [1 \, \, e^{-i\omega t} \, \, e^{-i2 \omega t}   \dots \, \, e^{-iN\omega t}]$.

We can notice that if a vector $| a(\omega) \rangle $ does not belong to the noise subspace the denominator of $J(\omega)$ tends to 0. As a consequence, the frequencies that span the signal subspace will correspond to the peaks of the $J(\omega)$ function. Moreover, we also highlight that, given the dimensionality of the signal, the resolution of the imaging function can be arbitrarily high. This effect is based on the superresolution property of the MUSIC algorithm, i.e., the capability to resolve two peaks separated below 1 Rayleigh length\cite{liao2016music}.

Finally, we note that this formulation of the MUSIC algorithm allows one to extract frequencies from a single snapshot of the signal in place of multiple snapshots as per the original formulation of the algorithm\cite{schmidt1986multiple}. We refer to Sec.\,\ref{MUSIC properties} for a summary of the main formal results and properties of this protocol.

We now conclude the theoretical section of this work, having discussed the main steps of the proposed algorithm (see also Fig.\,\ref{algorithm_summary}): (i)~a quantum computer is used to sample a randomly chosen subset of the autocorrelation function, generated via a Hadamard test. (ii)~The sampled points, along with an initial guess for the underlying signal, are used to reconstruct the full autocorrelation function (Algorithms~\ref{IVDST}--\ref{initialization}). (iii)~Finally, the dominant frequencies in the recovered signal are extracted by identifying the maxima of the imaging function (Eq.\,\ref{imaging_function_eq}) computed using the MUSIC algorithm. 
In the next section, we present the results of our implementation, specifically applied to quantum chemistry Hamiltonians.

\section{Numerical results}\label{numerics}

In this section, we show the results obtained from the algorithm implementation discussed in the previous section and summarized in Fig.\ref{algorithm_summary}. First, we give the technical information needed to reproduce the results shown in Sec.\,\ref{heisenberg_scaling}-\ref{input_state_sec}. 
Our numerical analysis is aimed at demonstrating three fundamental points: (i) the protocol is able to simultaneously estimate multiple eigenvalues with chemical accuracy while maintaining Heisenberg scaling (see Sec. \ref{heisenberg_scaling}); (ii) the algorithm is able to obtain accurate results even when the spectrum is very dense (see Fig.\,\ref{music_resolution}) and (iii) it allows us to study systems for which we cannot classically prepare initial wave functions with high overlaps (see Fig.\,7).

\subsection{Computational details}\label{comp_details}

All the results shown in this section have been obtained with a code distributed open-source in the following GitHub repository\cite{compressed_qpe_code}. The Jordan-Wigner mapping was used in all our calculations, as implemented in the PennyLane library\cite{bergholm2018pennylane}. The initial input wave function has been prepared and uploaded to the quantum register with a custom modification of the Overlapper library\cite{Overlapper}. The classical emulation of the Hadamard circuit has been performed with a homemade JAX code\cite{jax2018github}. Finally, the modified IVDST and MUSIC algorithm have been implemented using JAX and the NumPy\cite{harris2020array} library. The implementation of the MUSIC algorithm has been partly taken from Ref.\,\cite{musicimpl}.

If not stated differently, the input state wavefunctions have been prepared picking, for each desired eigenstate, the determinants with an overlap greater than 0.1 w.r.t. a CISD reference wavefunction.

The IVDST algorithm has always been run setting a convergence threshold of $10^{-6}$ or a maximum number of iterations $n_{it} = 1000$. The learning rate for the gradient descent step is $\delta = 0.5$ for the first iteration and $\delta = 0.01$ for the remaining optimization. The thresholding of the Toeplitz matrix has been performed with a value of $\tau = 0.001$ (step 5 of Alg.\,\ref{IVDST}). As also mentioned in the work of Wang \textit{et al.}\cite{wang2018ivdst}, the choice of $\delta$ and $\tau$ is made by balancing the accuracy and speed of the calculation. Here we set the parameters to ensure a quite accurate reconstruction of the signal; nonetheless, we point out that the full reconstruction for a signal of 1000 points with these parameters takes approximately 5 minutes on a standard desktop computer. The initial guess was computed according to Alg.\ref{og_initialization} and we set the initial guess coefficients all equal to $c = \frac{1}{\sqrt{s^*}}$ where $s^*$ denotes the number of frequencies targeted by the calculation. The initial guess frequencies were chosen to have an error of around $0.1-0.01 \,\,\text{Ha}$ w.r.t. the exact frequencies. We made this choice to resemble the situation in which our initial estimate is only a rough approximation of the true frequency. 

In all our calculations, the number of samples is set to $\text{\#}_{\text{samples}} = s \, \log s \, \log N$, where $s$ denotes the estimated number of frequencies with a non-negligible contribution to the signal. We then set the number of shots per sample to $\text{\#}_{\text{shots}} = \sqrt{s \, \log s \, N \, \log N}$. 

We remark that the number of sampled points is determined by the main results of compressed sensing theory, as reported in Sec.\,\ref{recovery_section}. On the other hand, the number of shots per sample can be estimated using the results in \cite{ding2023even} (Appendix A), where the authors showed that in order to maintain the noise amplitude $|\xi(\eta)| < \sigma_H$ (with $\sigma_H$ being a noise tolerance parameter) with probability at least $1 - \delta$, it is sufficient to acquire a number of shots on the order of $\mathcal{O} (\log(\text{\#}_{\text{samples}}) \, \delta / \sigma_H^2)$. 

We could have based our settings on this bound but we decided to opt for a more practical definition, expressed directly in terms of the total signal length and the number of dominant frequencies. In the following we will come back to this point to understand how our choice relates to the bounds of Ref.\,\cite{ding2023even}.

Frequency detection was performed using the peak finding routine as implemented in the SciPy package\cite{2020SciPy-NMeth} on the imaging functions $J(E)$ obtained with the MUSIC algorithm.

Most of the results presented in the next sections show plots of the average results of the compressed QPE protocol as a function of the total execution time (Eq.\,\ref{runtime}) or the maximum evolution time (between sampled points) $T_{max}$. For this purpose, we performed several executions of the compressed QPE protocol. This is necessary to discuss an algorithm that is inherently based on a random initial guess (in this case the selection of points to be measured). The results in Fig.\,\,\ref{heinsenb-scaling1}-\ref{heinsenb-scaling2} were obtained by averaging the results obtained with the compressed QPE protocol five times for each fixed signal length calculation. In addition, we scanned over the signal length with a step size of 50 points. The final results were obtained by binning the total runtime axis (or $T_{max}$ axis) with 20 bins and calculating the mean and standard deviations within a given bin.

The results shown in Fig.\,7b-c have been obtained with a similar procedure. For each fixed signal length calculation we repeated the compressed QPE calculation with three different randomly drawn initial samples. Furthermore, we scanned, for a fixed value of number of determinants used for the initial input state, over the signal length from 400 points to 900 points with a step size of 100 points. In total, the results shown in Fig.\,7b-c were obtained by performing 285 independent calculations. For each value of the number of determinants used, 15 different calculations were carried out, differing either in the initial drawn samples or in the total signal length.

\subsection{Heisenberg scaling in the weak and strong correlation regime}\label{heisenberg_scaling}

\begin{figure*}[t!]
    \centering
    \begin{tikzpicture}
        
   \node[inner sep=0pt] (img) at (0,0) {\includegraphics[width=0.75\textwidth]{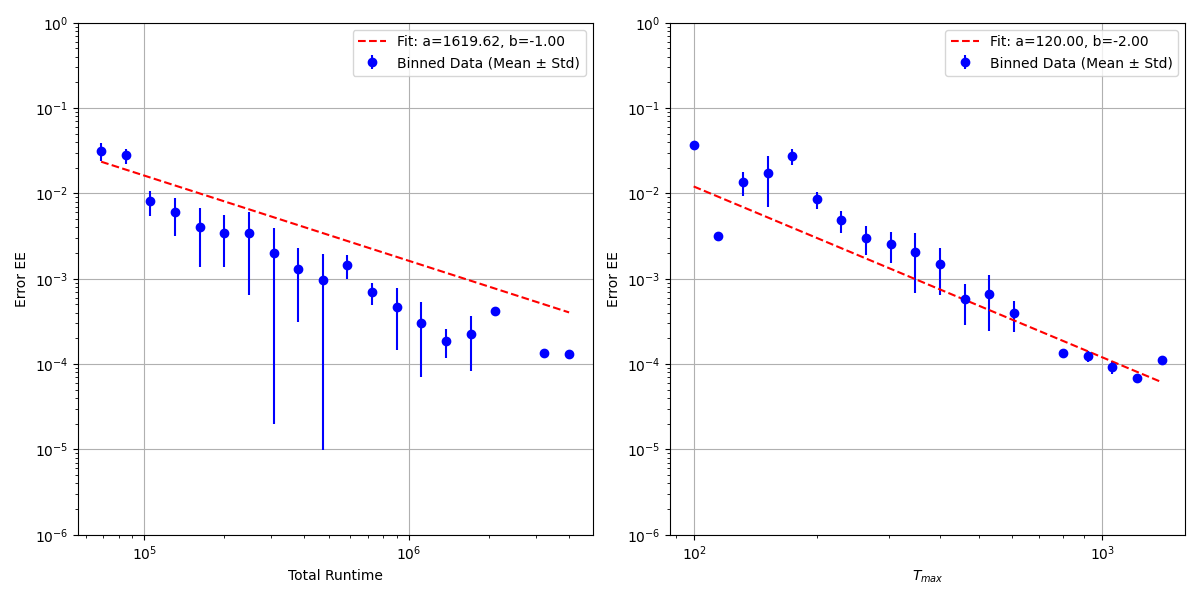}};

           \node[font=\small, xshift=5pt, yshift=-5pt] at (-5.75,3) {(a)};
        \node[font=\small, xshift=5pt, yshift=-5pt] at (.85,3) {(b)};
        \node[font=\tiny, xshift=0.5pt, yshift=-2.5pt, rotate=90] at (-6.5, 1) {[Ha]};
        \node[font=\tiny, xshift=0.5pt, yshift=-2.5pt, rotate=90] at (.15, 1) {[Ha]};
        \end{tikzpicture}
    \caption{Heisenberg-limited scaling in the weak correlation regime. Estimation of four different eigenfrequencies ($|GS\rangle, |S_1\rangle, |S_4\rangle \,\, \text{and} \,\, |S_5\rangle$) for the LiH molecule at equilibrium geometry. Red dashed lines shows a power-law dependency ($y = a x^{-b}$) of the errors obtained from the compressed-QPE calculations (blue dots) as a function of the total runtime (a, see Eq.\ref{runtime}) and of the maximal evolution time sampled with the Hadamard test (b). Blue bars represent standard deviation for the different runs of calculations performed (i.e. averaging of different randomly sampled points) and binning different Compressed QPE runs as explained in Sec.\,\ref{comp_details}.}
    \label{heinsenb-scaling1}
\end{figure*}

As discussed above, the objective of this work is to propose a new algorithm for the simultaneous estimation of multiple eigenvalues of a quantum Hamiltonian $H$.

Given the great interest of the scientific community in the possible impact that quantum computation may have in the field of computational chemistry, in this work we focus on the numerical analysis of the new algorithm specifically applied to the molecular Hamiltonian $H_{mol}$:

\begin{equation}
\label{molecular_hamiltonian}
    H_{mol} = \sum_{p,q} h_{pq} a^{\dagger}_{p}a_q + \frac{1}{2}\sum_{p,q,r,s} g_{pqrs} a^{\dagger}_{p}a^{\dagger}_{r}a_{s}a_{q}
\end{equation}
Where $a_p$ ($a^{\dagger}_p$) is  the annihilation (creation) operator acting on the spin-orbital $p$, $h_{pq}$ are the one-electron integrals containing the kinetic energy and the electron-nuclei repulsion terms and $g_{pqrs}$ are the two-electron repulsion integrals.

To assess the potential impact of our algorithm on quantum chemistry, it is crucial to clarify the terms under which a quantum algorithm is expected to influence this field~\cite{motta2022emerging}. Broadly speaking, the challenge of classically simulating a molecular system depends on its degree of compactness, that is, the number of electronic configurations needed to accurately describe its quantum state~\cite{mcclean2015compact}. The classification of electronic correlation in molecular systems has been explored extensively in the literature~\cite{morchen2024classification, cioslowski2012robust, materia2024quantum}. For the purposes of this work, it is sufficient to understand how this phenomenon affects correlation functions.

As pointed out by Zurek~\cite{zurek1982environment}, both the decaying speed and the amplitude of fluctuations around the long time average in the correlation functions depend on the number of states involved in the dynamics. In particular, as the number of contributing states increases, the autocorrelation function tends to decay more rapidly towards its long time average. This behavior can also be interpreted in terms of the orthogonality catastrophe\cite{anderson1967infrared}: the evolution of the system more readily tends toward a state whose average overlap with any eigenstate of the system decreases exponentially with system size. 
On the other hand, as the number of contributing states increases (and thus the number of terms included in $z_{nr}$ in Eq.\,\ref{relevant_not_relevant}) the fluctuations around the long-time average value will decrease.

These two combined effects can affect the efficiency of our protocol because both influence the relative weight that $z_r$, $z_{nr}$ and $\xi(\eta)$ have in the autocorrelation function, respectively. Based on these considerations, the autocorrelation function for a weakly correlated system will exhibit oscillatory behavior in the long-time limit, even when the initial state consists of only a few determinants. In contrast, strong correlation leads to a rapid damping of the oscillations and reduces the amplitude of recurrence beatings. This damping is also related to the influence of noise: as the signal becomes less oscillatory, noise has a greater impact on the ability to extract meaningful information. In other words, the presence of many frequencies means that each individual frequency contributes only with a smaller component compared to the weakly correlated case. Moreover, the components tend to have more similar weights in the signal decomposition, making them comparable in magnitude to the noise, which further complicates information extraction.

We now discuss the results shown in Fig.\,\ref{heinsenb-scaling1}. Here we consider the LiH molecule at its equilibrium geometry. The choice of this system has been dictated by the fact that it has been largely used in the context of quantum computation as a prototypical example of a weakly correlated heteroatomic system\cite{castaldo2024fast, chan2021molecular, avramidis2024ground}. In this first example we try to estimate \emph{simultaneously} four different eigenvalues corresponding to the ground state, the first, the fourth and the fifth singlet excited states. For this system the initial input state chosen (see Sec.\,\ref{comp_details}) is such that $|\langle\Psi_{input}| \Psi\rangle|^2 \approx 1$. 
First, as shown in Fig.\,\ref{heinsenb-scaling1}, we report the average error on the excitation energy as a function of the total runtime (panel \ref{heinsenb-scaling1}a) and of the maximal time at which the autocorrelation function has been measured ($T_{max}$, panel \ref{heinsenb-scaling1}b). Regarding the latter it is important to notice that $T_{max}$ does not exactly correspond to the total length $N$ of the signal vector for two reasons: (i) in our processing routine we have randomly sampled the wavefunctions on the positive axis of time but we then reconstructed the autocorrelation function in the time domain interval $[-T_{tot}, T_{tot}]$ doubling the total number of points; (ii) the random drawing prescribed by the compressed sensing protocol is such that in general $T_{max} \leq T_{tot}$.

As a first result, we note that the protocol is able to recover all the four frequencies within chemical accuracy, i.e. $\epsilon \leq 1\,\,mHa$. Then, looking at Fig.\,\ref{heinsenb-scaling1}a, we note that error estimates (blue dots) closely follow the red dashed line posing the Heisenberg scaling (i.e., $\epsilon \propto \mathcal{O}(T_{runtime}^{-1})$). This is an important result, since -- as mentioned in the introduction -- only a few other algorithms have demonstrated such (optimal) scaling for the general case of multiple eigenvalue estimation \cite{ding2023simultaneous, ding2024quantum}.

We now examine the dependence of the average error on the excitation energy (EE) as a function of \( T_{\text{max}} \). Two main effects are noteworthy:

\begin{itemize}
    \item For lower values of \( T_{\text{max}} \), the error scaling resembles that of QPE, i.e., \( T_{\text{max}} \pi = \mathcal{O}(\epsilon^{-1}) \). However, as \( T_{\text{max}} \) increases, this regime is surpassed, and we observe performance comparable to other state-of-the-art algorithms, where \( T_{\text{max}} \pi \ll \epsilon^{-1} \). Reaching this regime is crucial, as it enables shallower circuits by reducing the maximum evolution duration.

    \item Additionally, we highlight the observed error dependence on \( T_{\text{max}} \), specifically \( \epsilon = \mathcal{O}(T_{\text{max}}^{-2}) \).
\end{itemize}
In commenting on this result, it is important to recall what mentioned earlier: for our method, $T_{\text{max}} \leq T_{\text{tot}}$, whereas in standard QPE, where $T_{\text{max}} = T_{\text{tot}}$, the scaling typically follows $\epsilon = \mathcal{O}(T_{\text{max}}^{-1})$. 

Here, we have chosen to report $T_{\text{max}}$ as the longest evolution time simulated on the quantum computer, since it is a metric that more accurately reflects the actual computational effort required by the algorithm.

However, we can also discuss the dependence of $\epsilon$ as a function of $T_{\text{tot}}$ (or equivalently, the number of steps $N$). In particular, as shown in \cite{liao2016music} and detailed in Appendix~\ref{MUSIC properties}, the scaling behavior of the MUSIC algorithm allows to outperform, even considering this metric, standard QPE and other methods found in the literature. An exception is the ESPRIT algorithm~\cite{li2020super}, a subspace method similar to MUSIC, which, as recently demonstrated\,\cite{ding2024esprit}, also exhibits an error scaling of $\epsilon = \mathcal{O}(T_{\text{tot}}^{-1.5})$.
To summarize, the results of Fig.\ref{heinsenb-scaling1} show that our algorithm achieves Heisenberg limit in the task of multiple eigenvalue estimation and that the error scales as $\mathcal{O}(T_{\text{max}}^{-2})$ according to the numerical experiments and $\epsilon = \mathcal{O}(T_{\text{tot}}^{-1.5})$ according to the analytical error bounds\,\cite{liao2016music}.

We now turn to analyze the performances of the algorithm in the strong correlation limit. We will consider the $H_8$ hydrogen chain which, although experimentally unstable\cite{stella2011strong}, have been extensively characterized as the simplest systems revealing strong electronic correlation phenomena\cite{motta2017towards, motta2020ground}. Besides this system having a strong multireference character, it also features a dense spectrum in the low energy section. This is an additional challenge for our routine as we need a resolution high enough to distinguish accurately between the different frequencies. The initial input state used for this calculations has been obtained uploading the first 120 most relevant determinants, from the exact FCI wavefunction, for each eigenstate that has been targeted. In this case we are trying to estimate the electronic energies of the ground state, the second, the fourth and the sixth singlet excited states. With this setting the overlap of the initial input state with the target eigenstates $\sum_{i \in \{GS, S_2, S_4, S_6 \} }|\langle \Psi_{input} | i\rangle |^2 \approx 0.55$.

\begin{figure*}[t!]
    \centering
    \begin{tikzpicture}
        \node[inner sep=0pt] (img) at (0,0) {\includegraphics[width=0.75\textwidth]{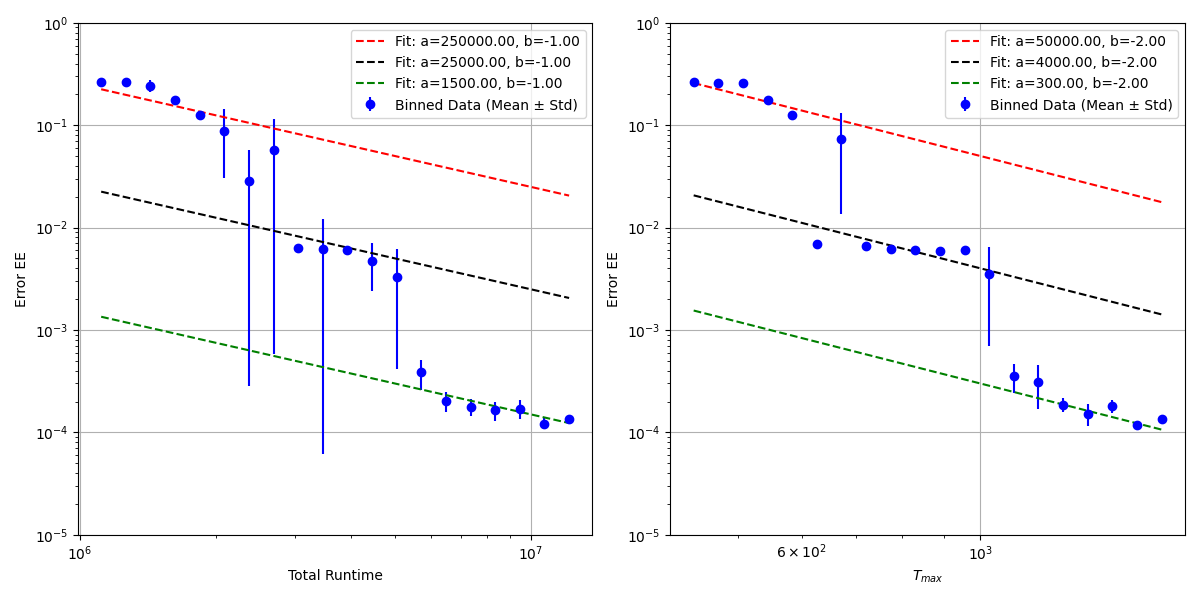}};
        \node[font=\small, xshift=5pt, yshift=-5pt] at (-5.75,3) {(a)};
        \node[font=\small, xshift=5pt, yshift=-5pt] at (.85,3) {(b)};
        \node[font=\tiny, xshift=0.5pt, yshift=-2.5pt, rotate=90] at (-6.5, 1) {[Ha]};
        \node[font=\tiny, xshift=0.5pt, yshift=-2.5pt, rotate=90] at (.15, 1) {[Ha]};

    \end{tikzpicture}
    \caption{Heisenberg-limited scaling in the strong correlation regime. Estimation of four different eigenfrequencies ($|GS\rangle, |S_2\rangle, |S_4\rangle \,\, \text{and} \,\, |S_6\rangle$) for the $H_{8}$ molecule at interatomic distance $r_{HH} = 2  \,\text{\r{A}} $. Dashed lines show a power-law dependency ($y = a x^{-b}$) of the errors obtained from the compressed-QPE calculations (blue dots) as a function of the total runtime (a, see Eq.\ref{runtime}) and of the maximal evolution time sampled with the Hadamard test (b). Blue bars represent standard deviation for the different runs of calculations performed (i.e. averaging over different randomly sampled points) and binning different Compressed QPE runs as explained in Sec.\,\ref{comp_details}. Different colors of the dashed lines refer to different resolution regimes allowed by the MUSIC algorithm, see Fig.\,\ref{music_resolution}.}
    \label{heinsenb-scaling2}
\end{figure*}

\begin{figure}[htbp!]
    \centering
    \includegraphics[width=\linewidth]{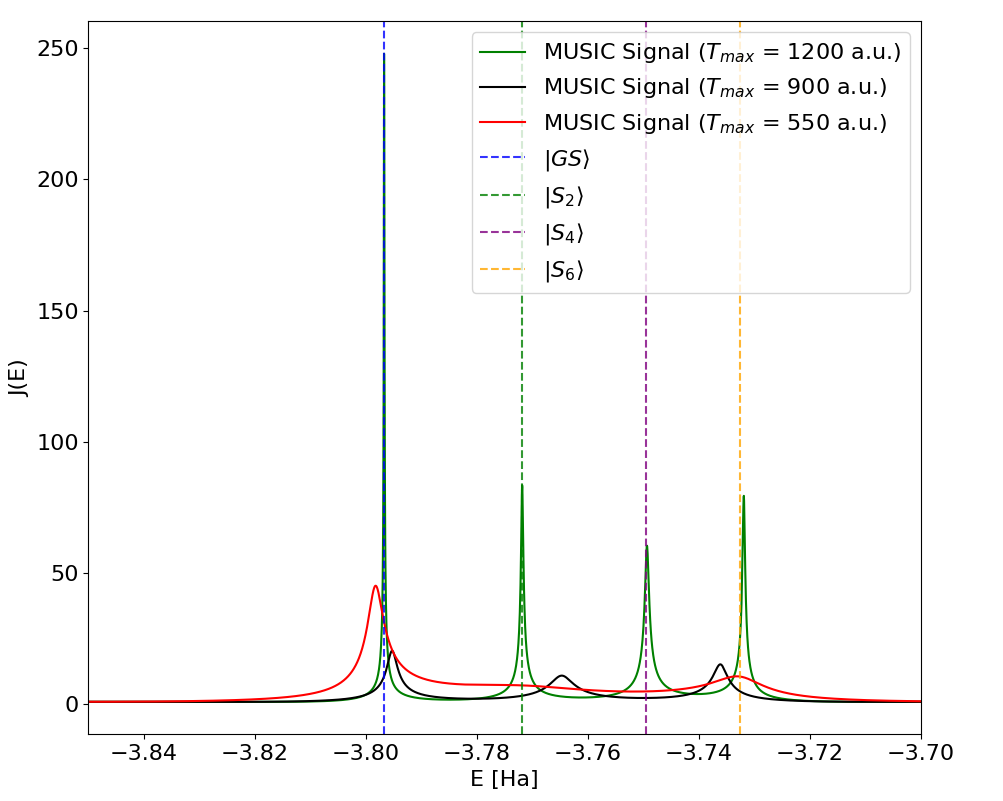}
    \caption{Resolution power of the MUSIC algoritm. Plot of the imaging function $J(E)$ (see Eq.\,\ref{imaging_function_eq}) obtained with signals of different length (solid lines). As a reference we also report the FCI energies (dashed lines) corresponding to the four different eigenvalues estimated in the Compressed QPE calculation.}
    \label{music_resolution}
\end{figure}

As we can see in Fig.\,\ref{heinsenb-scaling2}a, the power-law fit at Heisenberg scaling is still in good agreement with the errors obtained with the Compressed QPE but the overall picture is more complicated. In particular, we found that scanning over a range of calculations at increased runtime the algorithm errors follow three different regimes. To better understand the origin of this behavior in Fig.\,\ref{music_resolution} we compare the imaging functions obtained with the MUSIC algorithm for three different reconstructed signals of increasing length. As we can see, by increasing the $T_{max}$, hence the length of the signal, the algorithm is able to spot an increasing number of frequencies. We can rationalize this finding in terms of the analytical bounds reported in Sec.\,\ref{MUSIC properties} and of resolution properties of off-grid compressed sensing. First of all, as reported in Sec.\,\ref{MUSIC properties} the first frequency detectability condition derived in Ref.\,\cite{liao2016music} implies that the weakest detectable frequency is associated to a singular value $\sigma_s$ of the matrix $\textbf{W}$ which satisfies $\sigma_s \gg 2||E||_2$ with $E=Hank(z_{noise}(t))$. As mentioned in Sec.\,\ref{comp_details}, the number of shots per sample used in our scan is $\text{\#}_{\text{shots}} = \sqrt{s \, \log s \, N \, \log N}$ which means that the sampling noise component of the signal $|z_{noise}(t)|^2$ decreases approximatively as $\mathcal{O}(\frac{1}{\sqrt{N}})$ using the bound computed in Ref.\,\cite{ding2023even} (see also Sec.\,\ref{comp_details}). Moreover as mentioned in Sec.\,\ref{recovery_section}, as a consequence of reconstructing the signal with an off-the-grid method\cite{tang2013compressed}, we have a lower bound $\Delta_f \geq \frac{1}{N}$  on the minimal frequency gap in the spectrum of the underlying signal that can be precisely reconstructed.

These two facts allow us to rationalize the observed behavior of Figg.\,\ref{heinsenb-scaling2}-\ref{music_resolution}. This also means that for this system only the last points correspond to total runtime and $T_{max}$ values that satisfy both the resolution condition of the compressed sensing protocol and the frequency detectability condition of the MUSIC algorithm. From a computational point of view we can interpret these two conditions as a, system dependent, constant overhead of the algorithm. Focusing on these points we can see that all the conclusions drawn from the performance analysis on the weakly correlated limit still hold: the average error on the excitation energy decreases in good agreement with the Heisenberg scaling and with $\epsilon = \mathcal{O}(T^{-2}_{max})$.  

\begin{figure*}[htbp!]
\label{compressing_factor}
    \centering
    \begin{minipage}{0.45\textwidth}
        \centering
    \begin{tikzpicture}
        
      \node[inner sep=0pt] (img1) at (0,0)  {\includegraphics[width=\textwidth]{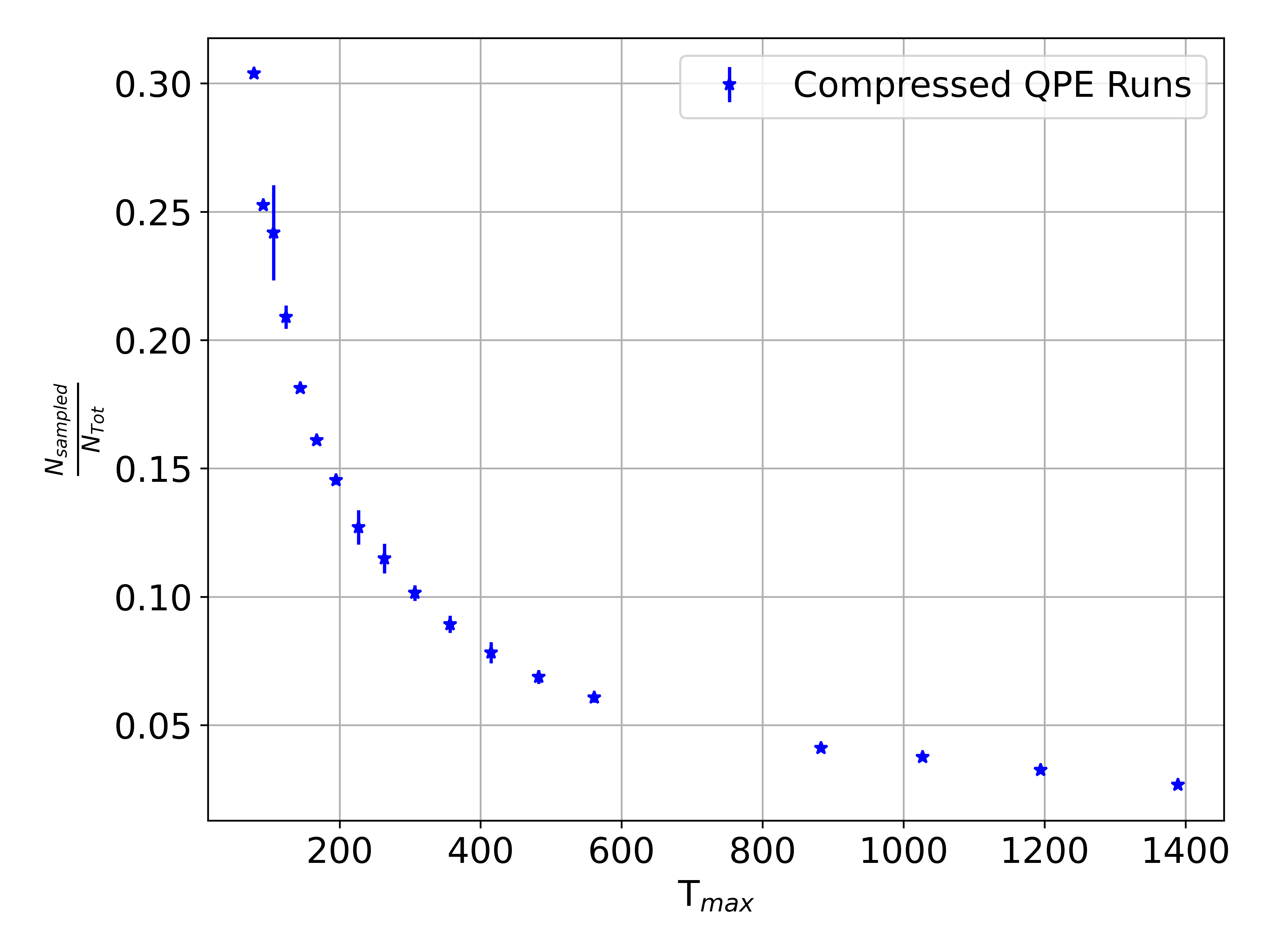}};
      \node[font=\small, xshift=0.5pt, yshift=-2.5pt] at (-1.75, 2.5) {(a)};

    \end{tikzpicture}

    \end{minipage}
    \begin{minipage}{0.45\textwidth}
        \centering
    \begin{tikzpicture}
        
    \node[inner sep=0pt] (img1) at (0,0)    {\includegraphics[width=\textwidth]{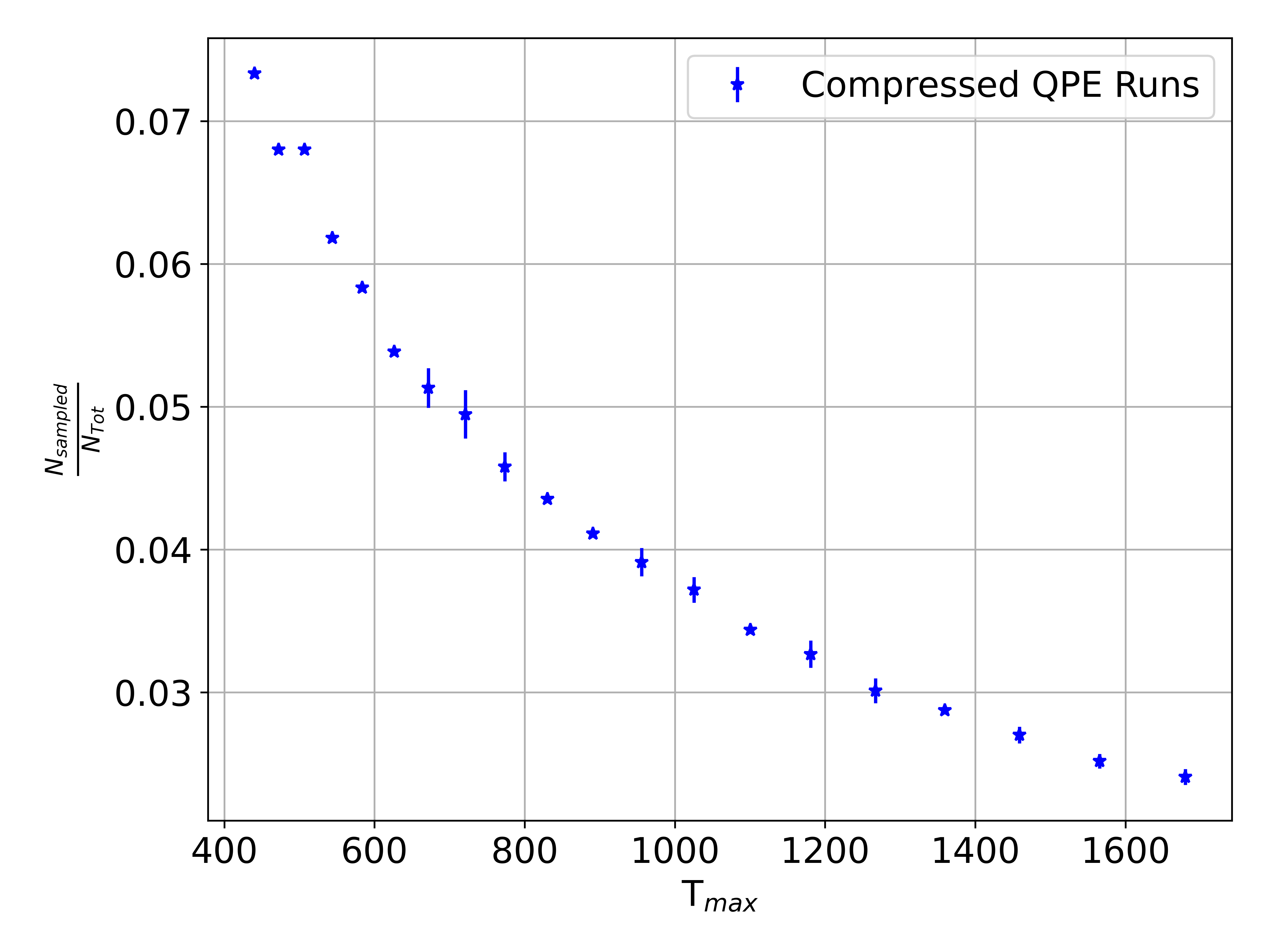}};
    \node[font=\small, xshift=0.5pt, yshift=-2.5pt] at (-1.75, 2.5) {(b)};

    \end{tikzpicture}

    \end{minipage}
    \caption{Computational saving due to the compressed sensing protocol. (a) Compressing factor as a function of the maximal evolution time sampled with the Hadamard test for the LiH molecule (see Fig.\,\ref{heinsenb-scaling1}). (b) Compressing factor as a function of the maximal evolution time sampled with the Hadamard test for the $H_8$ molecule (see Fig.\,\ref{heinsenb-scaling2}). Error bars are obtained averaging over multiple runs for the same calculation (i.e. different random sampling of the measurements) and binning different Compressed QPE runs as explained in Sec.\,\ref{comp_details}.}
\end{figure*}

To conclude this section, in Fig.\,6, we also report the compression factors, $\frac{N_{\text{sampled}}}{N_{\text{tot}}}$, as a function of $T_{\text{max}}$, obtained from the calculations performed for the LiH molecule in Fig.\,\ref{compressing_factor}a, and for the $H_8$ molecule in Fig.\,6b, respectively.

As we can see, even for a system like $H_8$ in the bond dissociation limit, where we used only 120 determinants out of 12{,}870 possible electronic configurations for each eigenstate, we were able to simultaneously estimate four different eigenvalues within chemical accuracy by sampling less than 5\% of the total number of reconstructed points.

\subsection{Effect of the initial input state}
\label{input_state_sec}

The question of finding and preparing good initial input states is closely related to the computational complexity of the electronic structure problem. The problem of finding ground state energies of local Hamiltonians in full generality is QMA-hard\,\cite{kempe2006complexity, o2022intractability}. On the other hand, in terms of total cost, implementing a classical state on a quantum computer is generally much less expensive than executing the full energy estimation algorithm, regardless of the method used \cite{fomichev2024initial, kiss2025early}.
These two facts, together with the orthogonality catastrophe \cite{anderson1967infrared}, motivate the investigation of how the choice of initial state affects the algorithm’s performance. A practical quantum advantage will, indeed, lie in the ability to determine eigenvalues at chemical accuracy starting from classically obtained initial states that do not quantitatively capture the system's chemistry.

\begin{figure*}[htbp!]

\centering

\begin{minipage}{0.32\textwidth}
    \centering
    \begin{tikzpicture}
        \node[inner sep=0pt] (img1) at (0,0) {\includegraphics[width=\textwidth]{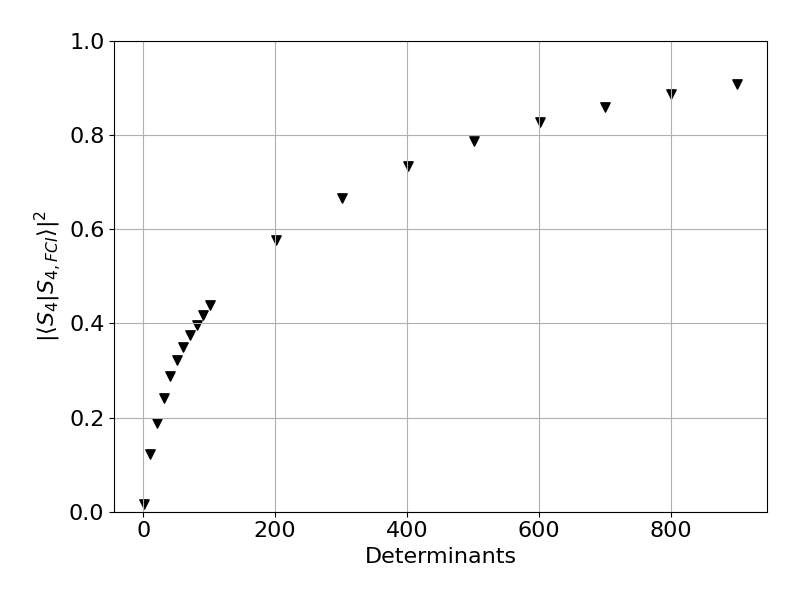}};
        \node[font=\small, xshift=0.5pt, yshift=-2.5pt] at (-1.5, 1.75) {(a)};
    \end{tikzpicture}
\end{minipage}
%
\begin{minipage}{0.32\textwidth}
    \centering
    \begin{tikzpicture}
        \node[inner sep=0pt] (img2) at (0,0) {\includegraphics[width=\textwidth]{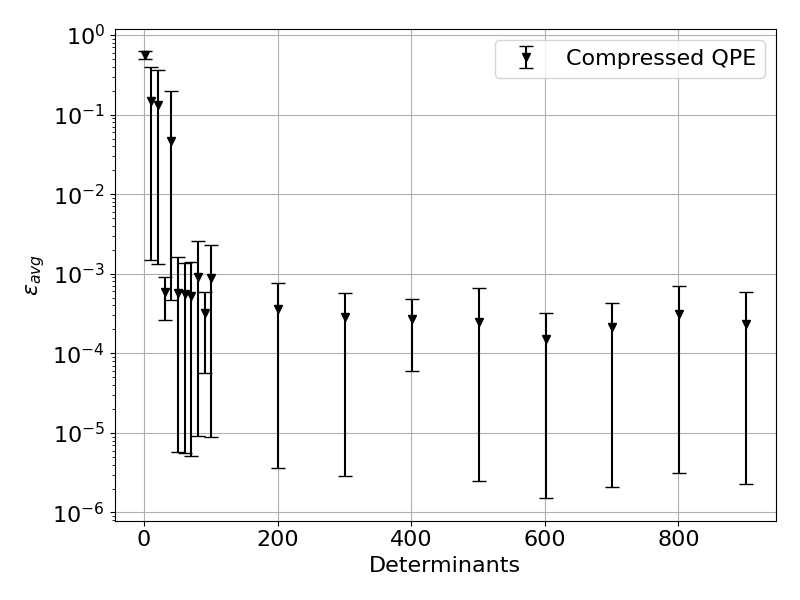}};
        \node[font=\small, xshift=0.5pt, yshift=-2.5pt] at (-1.5, 1.75) {(b)};
        \node[font=\tiny, xshift=0.5pt, yshift=-2.5pt, rotate=90] at (-2.65, .75) {[Ha]};

    \end{tikzpicture}
\end{minipage}
%
\begin{minipage}{0.32\textwidth}
    \centering
    \begin{tikzpicture}
        \node[inner sep=0pt] (img3) at (0,0) {\includegraphics[width=\textwidth]{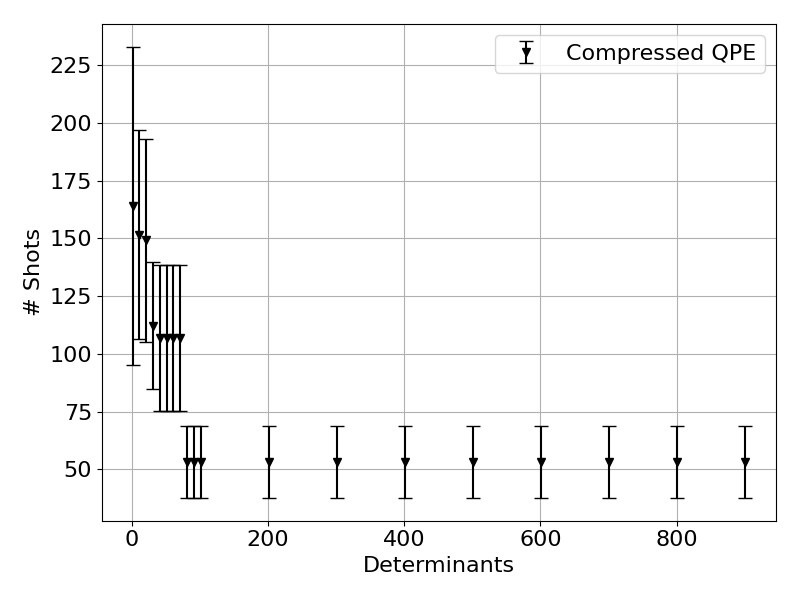}};
        \node[font=\small, xshift=0.5pt, yshift=-2.5pt] at (-1.5, 1.75) {(c)};
    \end{tikzpicture}
\end{minipage}
\label{initial_state_plot}
\caption{Effect of the initial input state on the energy estimate accuracy. (a) Overlap of the initial input state used for the Compressed QPE protocol with respect to the exact $|S_{4, FCI}\rangle$ 4th singlet excited state of the $H_8$ molecule as a function of the determinants included. (b) Average error estimate as a function of the number of determinants included in the initial input state; error bars are obtained averaging multiple runs for the same Compressed QPE protocol (i.e., different random sampling) and averaging over the results obtained varying the total autocorrelation function length, see also Sec.\ref{comp_details}. (c) Number of shots per sample acquired for the different Compressed QPE runs as a function of the number of determinants included in the initial input state; error bars are obtained as per Fig.\,7b\,.}
\end{figure*}
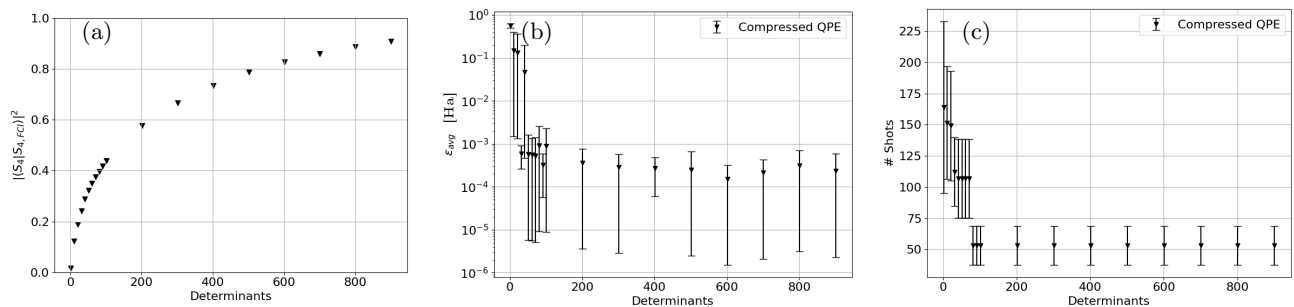

For all these reasons we decided to study the effect of the initial input state on the fourth excited singlet state of the $H_8$ molecule, Fig.\,7 . This case is particularly challenging as the strong multireference character is not only intrinsic in the system itself but is now enhanced by the fact that we are looking at an excited state built with a CI expansion which uses as a reference the HF determinant. Since the HF determinant is built to describe the system in its ground state, such an expansion will be even more dense for a highly excited state such as $|S_4\rangle$.

As we can see, the recovery algorithm is able to obtain estimates of the excited-state energy within chemical precision for overlap values $|\langle S_4|S_{4, FCI}\rangle|^2 \approx 0.3$. Finally, to enable a fair comparison across calculations using different initial input states, we adjust the total number of shots per sample accordingly. Specifically, reducing the number of determinants in the initial state increases the number of dominant frequencies contributing to the autocorrelation signal. As a result, the total number of shots required per sample also increases. Nevertheless, as shown in Fig.\,7c, even with the increased runtime, the accuracy of the protocol decreases when a poorer initial state is used.

These result suggests that for practical applications in quantum chemistry our method is a valuable option to study systems beyond the reach of classical computers. A thorough comparison of the performance of the main quantum algorithms for multiple eigenvalue estimation in the EFTQC scenario for quantum chemistry is beyond the scope of this study, we plan in a subsquent work to investigate the effect of the initial input state in realistic systems across various algorithms\cite{ding2023even, ding2024quantum, li2023adaptive, stroeks2022spectral}.




\section{Conclusions}

In this work we propose an algorithm to simultaneously estimate multiple eigenvalues of a molecular Hamiltonian by reconstructing, via an off-the-grid compressed sensing method, the full autocorrelation function from few measurements performed on a quantum computer. This work extends the results of \cite{yi2024quantum}, which had focused on the single quantum phase estimation task.

Our results show that the proposed method outperforms the standard QPE algorithm and is on-par with other state-of-the-art proposals for multiple eigenvalue estimation.

Given the importance that these methods might have in the chemical domain, we presented our implementation showing that the algorithm exhibits Heisenberg scaling even for systems whose electronic structure is very difficult to describe by conventional methods and whose spectrum is very dense (i.e., where high resolution is required to distinguish two eigenvalues with chemical accuracy). We remark that, despite the numerical evidences of Heisenberg scaling, an analytic bound that formally demonstrates it is still missing.
We have also characterised the algorithm not only in terms of the computational resources required to obtain an $\epsilon$-accurate estimate of the excitation energies, but also in terms of the maximum evolution duration that needs to be implemented on the quantum computer. This figure of merit is particularly important because it is related to the depth of the circuits that need to be implemented. 
In this respect, the systematic comparison, from a performance standpoint, of the most promising algorithms for EFTQC applied to chemical systems will be the objective of a subsequent study. This is in fact also the first study in which this type of algorithm is applied to chemical systems of remarkable complexity such as the excited states of $H_8$ in the bond dissociation limit and little can be said for the performance of other proposed algorithms w.r.t. these systems. As recently shown by Kiss et al.\,\cite{kiss2025early}, the practical implementation of early fault-tolerant algorithms can uncover subtleties that are not readily apparent when relying solely on an algorithmic perspective. Another important question to address is the following: under what conditions is it more advantageous to estimate a single eigenvalue independently rather than estimating K eigenvalues simultaneously? The discussion presented here on the role of the initial state provides a foundation for further investigation in future work.

As a further perspective for this work, we may consider the extension of this algorithm to reconstruct non-stationary signals: that is, signals whose coefficients are also time-dependent. Such an approach could reduce the overhead needed for the implementation of error-correction as we could include the effect of noise as a dampening of our signal to be reconstructed. Similar approaches based on the calculation of the Fisher information matrix\cite{bolzonello2024fisher} have already been proposed in the context of sub-Nyquist sampling for time-domain spectroscopy and could be an interesting avenue to explore forms of error mitigation in the field of quantum simulation and towards the application on current quantum hardwares.

Finally, another future perspective of this work is the extension to arbitrary correlation functions, in a similar spirit of Ref.\,\cite{zhang2022computing}, for the evaluation of a general molecular observable.
This last point looks into the direction of expanding the scope of quantum algorithms to target more complex systems. As already pointed out in Ref.\,\cite{capone2024vision} a long term vision for physical chemistry simulation entails the inclusion of quantum algorithms into a multiscale framework: whether current approaches for EFTQC can be efficiently incorporated into a multiscale model is still an open question.

%

\emph{Code availability $-$}
All the code needed to reproduce the results of this work is available open-source at\,\cite{compressed_qpe_code}.

\emph{Acknowledgments $-$}
The authors acknowledge the
University of Padova Strategic Research Infrastructure
Grant 2017: ‘CAPRI: Calcolo ad Alte Prestazioni per
la Ricerca e l’Innovazione’ and the WCRI Quantum Computing
and Simulation Center of Padua University for HPC usage. D.C. also
acknowledges usage of the ETH local computational facilities
for HPC usage.
The authors also acknowledge fundings from the MUR PNRR-PRIN2022 (PNRR M4C2
Inv 1.1) Grant 2022W9W423 ‘Quantum computing for
Computational Chemistry and Materials Science’ funded
by the European Union—NextGenerationEU
and from the NextGenerationEU PNRR project CN00000013 -
Italian Research Center on HPC, Big Data, and Quantum
Computing . 

%
%

\section*{Supplemental material}

\subsection{Properties of the MUSIC algorithm}
\label{MUSIC properties}

In this section, we report some properties of the MUSIC algorithm without their proofs. For a more detailed account and all the proofs we refer to Ref.\,\cite{liao2016music}.

\begin{lemma}
    Let $E$ be the Hankel matrix associated with the noisy component of $\Tilde{z}(t)$, $E=Hank(z_{noise}(t))$.Let $\sigma_1 > \sigma_2 > \dots \sigma_s > \sigma_{s+1} \dots \sigma_N$ denote the singular values of the Hankel matrix associated to $\Tilde{z}(t)$, $\textbf{W}=Hank(\Tilde{z}(t))$. A frequency belongs to the signal subspace $\mathcal{F}_s$ if

\begin{equation}
        \sigma_s \gg 2||E||_2.
\end{equation}
\end{lemma}




\begin{lemma2}
Let $\sigma_1 > \sigma_2 > \dots \sigma_s > \sigma_{s+1} \dots \sigma_N$ denote the singular values of the Hankel matrix associated to $\Tilde{z}(t)$, $\textbf{W}=Hank(\Tilde{z}(t))$. The maximum number of detectable frequencies is associated with a sharp drop in the singular value distribution of $H$, i.e. $s$ s.t. $\sigma_s \gg \sigma_j \, \forall j \geq s+1$.
\end{lemma2}

\begin{lemma3}
Let $\omega_j$ be the true frequency associated with the singular value $\sigma_j$ and $\hat{\omega}_j$ be the estimate provided by the MUSIC algorithm. Let $N$ be the total length of the recovered vector $\Tilde{z}(t)$. If $\sigma_j$ satisfies the frequency detectability condition then the error $\epsilon = |\omega_j -\hat{\omega}_j|$ is given by:

\begin{equation}
    \epsilon = |\omega_j -\hat{\omega}_j| = \mathcal{O}(\frac{log(N)}{N^{\frac{3}{2}}})  . 
\end{equation}
\end{lemma3}

\subsection{MUSIC denoising}

\begin{figure}[htbp]
    \centering
    \includegraphics[width=\linewidth]{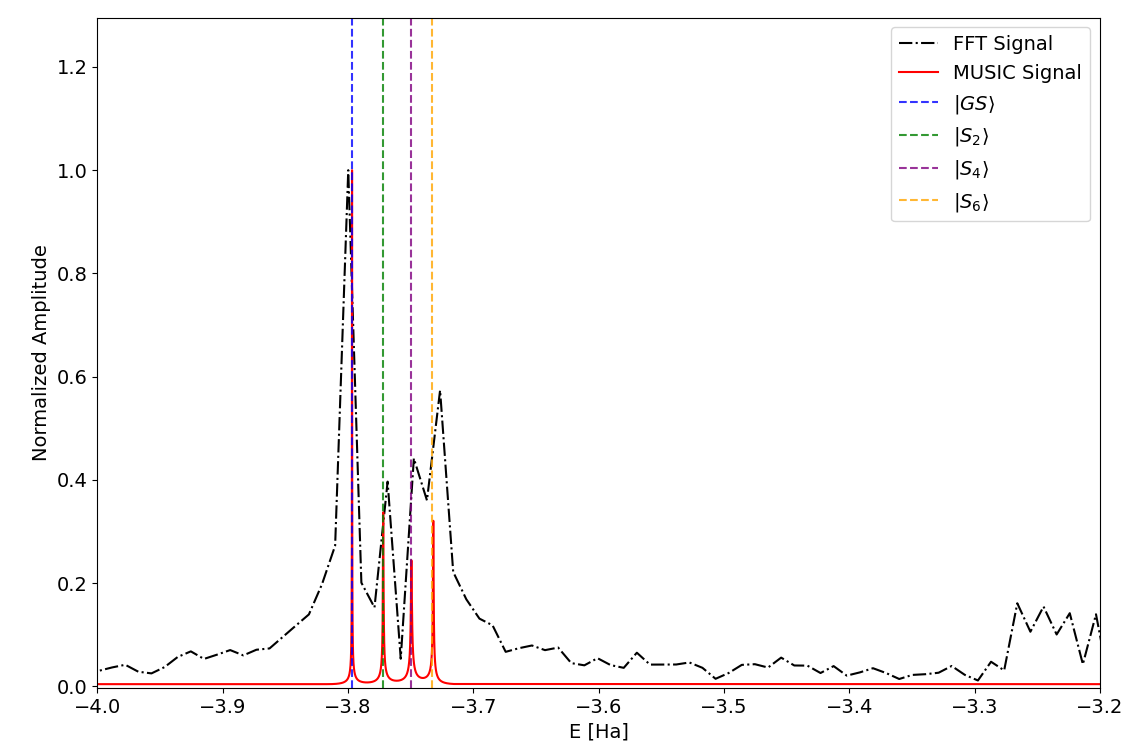}
    \caption{Comparison between FFT (black dashed-dotted line) and MUSIC (red solid line) transformed signals for a noisy autocorrelation function. Vertical dashed lines provide reference values for the exact energies of the eigenstates targeted by this calculation.}
    \label{fft2music}
\end{figure}

It is already well-known in literature that subspace methods like ESPRIT\cite{li2020super} or MUSIC\cite{liao2016music} outperform the accuracy estimates of the standard FFT when it comes to noisy signals. In this section we report examples of this fact for an example calculation of the ones discussed in the main text (see Sec.\ref{numerics}, Fig.\ref{heinsenb-scaling2}). Particularly, in Fig.\ref{fft2music} we plot both the MUSIC imaging function (solid line) and the spectrum obtained via direct Fourier transform of an autocorrelation function ($T_{max}=1200\,\ a.u.$) of the $H_8$ molecule. As we can see, even though the FFT is able to clearly identify the position of the dominant frequencies (vertical dashed lines give the reference FCI value), only the MUSIC algorithm is able to retrieve the frequencies within chemical accuracy (see Tab.\ref{tab:fft_music}). Interestingly, looking at the Fourier spectrum, we can also see a clear contribution coming from other excited states which the MUSIC algorithm completely suppresses as they belong to the noise subspace as we defined in all our calculations the dimension of the signal space to be equal to the number of eigenvalues that we wanted to target with the calculation.

\begin{table}[htbp]
    \centering
        \caption{Comparison of FFT and MUSIC energy error estimates. All values are reported in Hartree.}
    \begin{tabular}{l c c c c}
        \toprule
         & $\epsilon_{|GS\rangle}$ & $\epsilon_{|S_2\rangle}$ & $\epsilon_{|S_4\rangle}$ & $\epsilon_{|S_6\rangle}$ \\
        \midrule
        FFT   & 0.00305043 & 0.00357992 & 0.00203543 & 0.00612200 \\
        MUSIC & 2.599e-05  & 8.293e-05  & 9.701e-05  & 7.156e-04  \\
        \bottomrule
    \end{tabular}
    \label{tab:fft_music}
\end{table}

\subsection{Preconditioning the IVDST allows for faster and more accurate signal recovery}\label{preconditiong_results}

In Sec.\,\ref{recovery_section}, we introduced the recovery algorithm employed in this work and discussed the modifications made to the original proposal \cite{wang2018ivdst} to better suit the reconstruction of noisy signals, such as those obtained from a finite number of quantum measurements. We also highlighted that the initialization strategy presented in Alg.\,\ref{initialization} enables faster and more accurate signal recovery. In the main text, we presented a convergence plot from a representative calculation used in the numerical analysis of Sec.\,\ref{numerics}. Here, we further support our claims by showing results from multiple calculations on the LiH molecule, where we varied the length of the acquired and reconstructed signals. The results demonstrate that the physically informed initialization (blue dots) consistently yields more accurate estimates compared to the original initialization scheme of Alg.\,\ref{og_initialization} (red dots). We note that, unlike the calculations in Sec.\,\ref{numerics}, all simulations here were performed with an equal number of samples and shots per sample, i.e., $\text{\#}_{samples} = \text{\#}_{shots} = s\,logs\,logn$ .

\begin{figure*}[htbp]
\label{precond_effect_cqpe}

\begin{tikzpicture}[node distance=cm,
    every node/.style={fill=white, font=\sffamily}]
    
   \node (figure) at (0,0) { \centering
    \includegraphics[width=\textwidth]{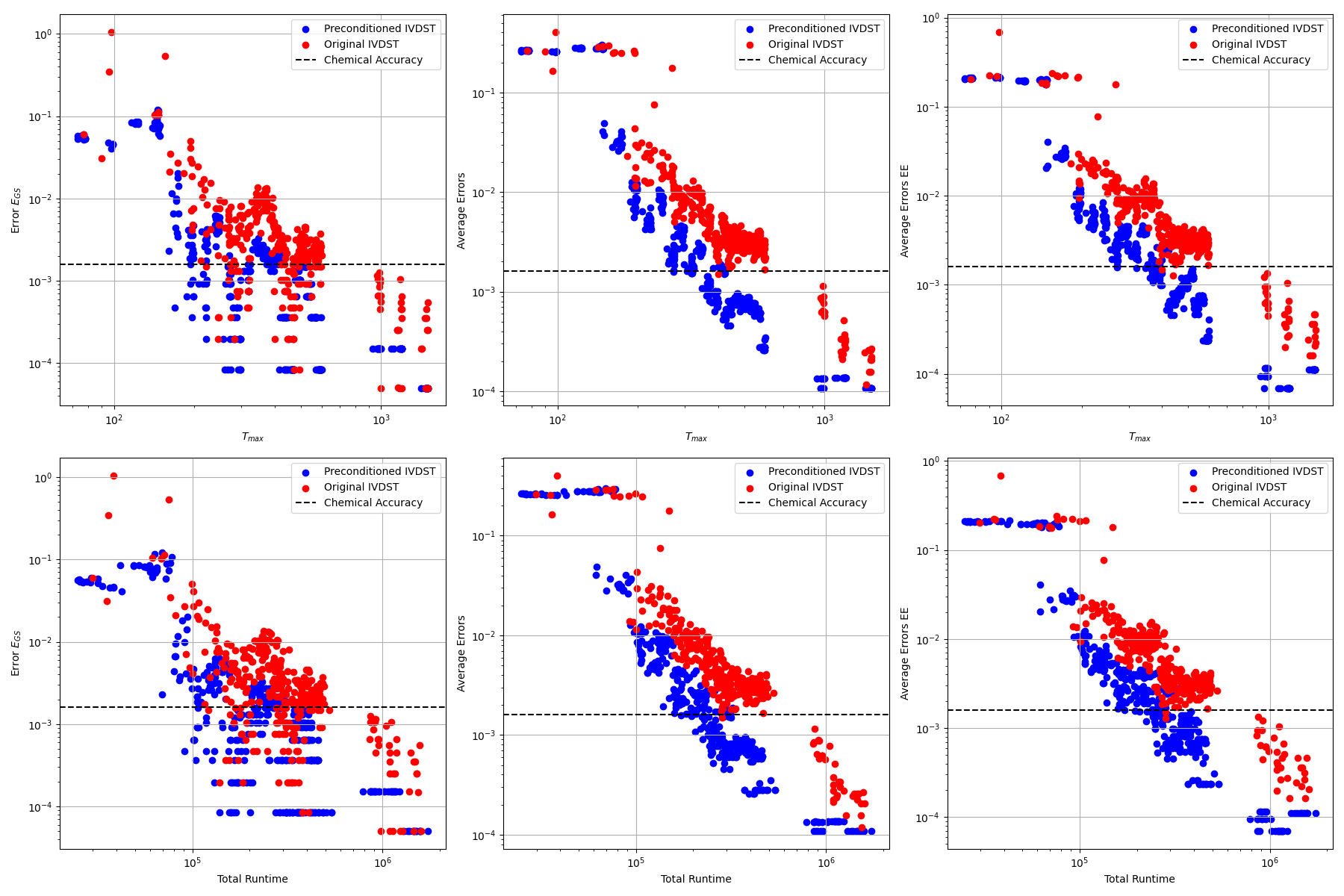}};

\node[font=\tiny, xshift=0.5pt, yshift=-2.5pt, rotate=90] at (-8.75, -2) {[Ha]};
\node[font=\tiny, xshift=0.5pt, yshift=-2.5pt, rotate=90] at (-8.75, 4) {[Ha]};
\node[font=\tiny, xshift=0.5pt, yshift=-2.5pt, rotate=90] at (-2.8, 4.15) {[Ha]};
\node[font=\tiny, xshift=0.5pt, yshift=-2.5pt, rotate=90] at (3.1, 4.2) {[Ha]};
\node[font=\tiny, xshift=0.5pt, yshift=-2.5pt, rotate=90] at (-2.8, -1.8) {[Ha]};
\node[font=\tiny, xshift=0.5pt, yshift=-2.5pt, rotate=90] at (3.1, -1.7) {[Ha]};

\node (a) at (-7.8, 5.5) {a)};

\node (b) at (0.5, 5.5) {b)};

\node (b) at (6.5, 5.5) {c)};

\node (c) at (-7.8, -0.5) {d)};

\node (d) at (0.5, -0.5) {e)};

\node (b) at (6.5, -0.5) {f)};

\end{tikzpicture}
    \caption{Preconditioning the IVDST with approximate solutions allows for more accurate energy estimates. (a) Error on the ground state estimate, (b) average error on the total estimate for the different eigenvalues and (c) average excitation energy error as a function of $T_{max}$. (d) Error on the ground state estimate, (e) average error on the total estimate for the different eigenvalues and (f) average excitation energy error as a function of the total runtime (see Eq.\,\ref{runtime}). Red dots are obtained using the original initialization scheme proposed in \cite{wang2018ivdst}, blue dots are obtained with the initialization scheme summarized in Alg.\,\ref{initialization}.}
    \label{heinsenb-scaling}

\end{figure*}
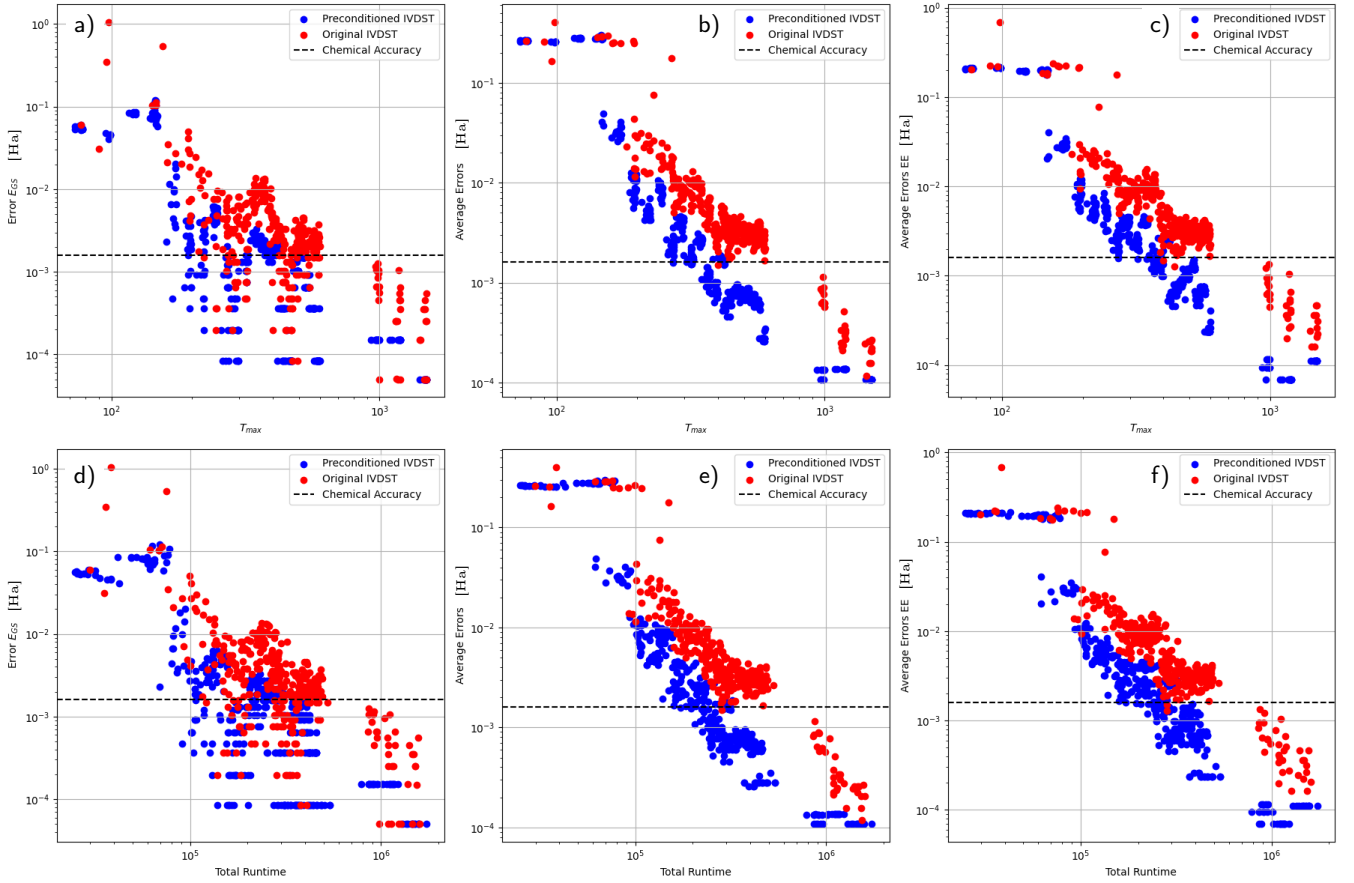

\bibliography{refs}


\end{document}